\documentclass[12pt]{article}
\usepackage{color}
\usepackage{amsmath} %Never write a paper without using amsmath for its many new commands 
\usepackage{amssymb} %Some extra symbols 
\usepackage[english]{babel}
\usepackage{graphics}
\usepackage[pdftex]{graphicx}
\usepackage{xspace}
\usepackage[gen]{eurosym}
\usepackage[font=footnotesize,format=plain,labelfont=bf,up,textfont=it,up]{caption}
\usepackage[left]{lineno}
\newcommand{\BARII}                   {1}
\newcommand{\BARIU}                 {2}
\newcommand{\BOLOGNAI}         {3}
\newcommand{\BOLOGNAU}       {4}
\newcommand{\CERN}                  {5}
\newcommand{\FRASCATI}          {6}
\newcommand{\GENOVAI}            {7}
\newcommand{\GENOVAU}          {8}
\newcommand{\GRANSASSO}     {9}
\newcommand{\KRACOW}          {10}
\newcommand{\KATOVICE}        {11}
\newcommand{\LECCEI}             {12}
\newcommand{\LECCEU}           {13}
\newcommand{\LECCEUU}        {14}
\newcommand{\LOSALAMOS}   {15}
\newcommand{\LOSANGELES} {16}
\newcommand{\MILANOB}          {17}
\newcommand{\MILANOP}          {18}
\newcommand{\MOSKOW}          {19}
\newcommand{\NAPOLII}            {20}
\newcommand{\NAPOLIU}          {21}
\newcommand{\PADOVAI}          {22}
\newcommand{\PADOVAU}        {23}
\newcommand{\PAVIAI}               {24}
\newcommand{\PAVIAU}             {25}
\newcommand{\ROMAU}            {26}
\newcommand{\WARSZAWAN}  {27}
\newcommand{\WARSZAWAT}  {28}
\newcommand{\NessieInstitutes}{
\BARII        . INFN, Sezione di Bari, 70126 Bari, Italy \\
\BARIU        . Dipartimento di Fisica dell'Universit\`a  di Bari, 70126 Bari, Italy \\
\BOLOGNAI     . INFN, Sezione di Bologna, 40127 Bologna, Italy \\
\BOLOGNAU     . Dipartimento di Fisica dell'Universit\`a  di Bologna, 40127 Bologna, Italy \\
\CERN             . CERN, Geneva, Switzerland\\
\FRASCATI    . Laboratori Nazionali di Frascati dell'INFN, 00044 Frascati (Roma), Italy \\
\GENOVAI        . INFN, Sezione di Genova, 16146 Genova, Italy \\
\GENOVAU        . Dipartimento di Fisica dell'Universit\`a di Genova, 16146 Genova, Italy \\
\GRANSASSO        . Laboratori Nazionali del Gran Sasso, INFN, 67010 Assergi (L'Aquila), Italy \\
\KATOVICE       . Institute of Physics, University of Silesia, Katowice, Poland\\
\KRACOW       . Henryk Niewodniczanski Institute of Nuclear Physics, Polish Academy Science, Krak—w, Poland\\
\LECCEI        . INFN, Sezione di Lecce, 73100 Lecce, Italy \\
\LECCEU        . Dipartimento di Matematica e Fisica dell'Universit\`a  del Salento, 73100 Lecce, Italy \\
\LECCEUU     . Dipartimento di Ingegneria dell'Innovazione dell'Universit\`a del Salento, 73100 Lecce, Italy\\
\LOSALAMOS   . Los Alamos National Laboratory, New Mexico, USA\\
\LOSANGELES . Department of Physics and Astronomy, University of California, Los Angeles, USA\\
\MILANOB       . INFN, Sezione di Milano Bicocca, Dipartimento di Fisica G. Occhialini, 20126 Milano, Italy\\
\MILANOP       . INFN, Sezione di Milano e Politecnico, 20133 Milano, Italy\\
\MOSKOW       . INR-RAS, Moscow, Russia\\
\NAPOLII          . INFN, Sezione di Napoli, 80126 Napoli, Italy\\
\NAPOLIU        . Dipartimento di Scienze Fisiche, Universit\`a Federico II, 80126 Napoli, Italy\\
\PADOVAI      . INFN, Sezione di Padova, 35131 Padova, Italy \\
\PADOVAU      . Dipartimento di Fisica e Astronomia dell'Universit\`a  di Padova, 35131 Padova, Italy \\
\PAVIAI      . INFN, Sezione di Pavia, 27100 Pavia, Italy \\
\PAVIAU     . Dipartimento di Fisica Nucleare e Teorica, Universit\`a di Pavia, 27100 Pavia, Italy\\
\ROMAU        . Dipartimento di Fisica dell'Universit\`a  di Roma ``La Sapienza" and INFN, 00185 Roma, Italy \\
\WARSZAWAN    . National Center for Nuclear Research, Warszawa, Poland\\
\WARSZAWAT    . Institute for Radioelectronics, Warsaw University of Technology, Warsaw, Poland\\
* Also at Centre de Recherche en Astronomie Astrophysique et Gophysique, Alger, Algeria\\
}
\newcommand{\NessieAuthorList}{
A.~Antonello$^{\GRANSASSO}$,
D.~Bagliani$^{\GENOVAU, \GENOVAI}$,
B.~Baibussinov$^{\PADOVAI}$,
H.~Bilokon$^{\FRASCATI}$,
F.~Boffelli$^{\PAVIAI}$,
M.~Bonesini$^{\MILANOB}$,
E.~Calligarich$^{\PAVIAI}$,
N.~Canci$^{\GRANSASSO}$,
S.~Centro$^{\PADOVAU, \PADOVAI}$,
A.~Cesana$^{\MILANOP}$
% ...
\\
}
\begin{document}

%\date{}
\includegraphics[width=0.15\textwidth]{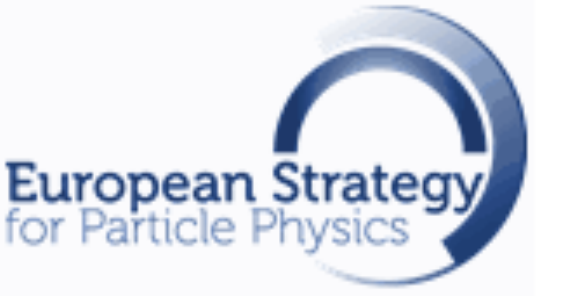}
\vskip -30pt \hspace{70pt}
\noindent{\em Contribution to the European Strategy for Particle Physics --}

\vskip -3pt\hspace{60pt} {\em  Open Symposium Preparatory Group, Kracow 10-12 September 2012}
%\title{\bf Search for anomalies in the neutrino sector with muon spectrometers and large LArÐTPC imaging detectors at CERN}
%\label{sec:future_exps_lar_nessie}

%\nopagebreak[4]

%\maketitle

\vskip 20pt
\begin{center}
{\bf\Large  Search for anomalies in the neutrino sector}
\vskip 7pt
{\bf\Large with muon spectrometers and}
\vskip 7pt
{\bf\Large large LArÐTPC imaging detectors at CERN}
\end{center}
\vskip 20pt

\vskip -20pt
\centerline{{\em ICARUS - NESSiE Collaborations}}
\vskip 20pt

%\author{\noindent \\ \NessieAuthorList }

\footnotesize{\noindent \\ \NessieAuthorList }

\begin{flushleft}
\footnotesize{\NessieInstitutes }
\end{flushleft}

\newpage

\subsubsection*{Introduction}

The long standing series of experimental results hinting at the hypothesis of {\em sterile neutrinos}~\cite{whitepaper} deserve 
in our opinion a major CERN investment to definitely clarify the underlying physics.

         Over the last several years, neutrinos have been the origin of an impressive
number of ``surprises''. Neutrino oscillations have so far established a beautiful
picture, consistent with the mixing of three physical neutrino $\nu_e$, $\nu_\mu$ and $\nu_\tau$ and
mass eigenstates $\nu_1$, $\nu_2$ and $\nu_3$. The two observed mass differences turn out to be relatively
small $\Delta m_{23}^2 \simeq 2.4 \times 10^{-3}$ eV$^2$ and 
$\Delta m_{21}^2 \simeq 8 \times 10^{-5}$ eV$^2$. The sum of the strengths of
the $\nu$'s has been found very near to 3. But it is possible that neutrinos are something
very different than just a neutral counterpart of charged leptons, leaving room for
additional neutrinos which do not see fully the ordinary electro-weak interactions but
still introduce mixing oscillations with ordinary neutrinos. Indeed there are a number
of ``anomalies'' which, provided they are confirmed experimentally, might be due to
the presence of larger squared mass differences related to additional neutrino states
with presumably some kind of ``sterile'' nature. Of course the astronomical
importance of neutrinos in space is immense, so is their role in the cosmic evolution.
Just a few eV neutrino mass may be the source of the dark mass.

       The possible presence of oscillations into sterile neutrinos was proposed by B.
Pontecorvo~\cite{pontecorvo}, but so far without conclusion. Two distinct classes of anomalies have
been reported, although not in an entirely conclusive level, namely:

\begin{itemize}
   \item  The observation of excess $\nu_e$ electrons originated by initial $\nu_\mu$ beam from
       accelerators (LNSD~\cite{lsnd} / MiniBooNE~\cite{larnessie_5}). At present the LSND experiment and the
       MiniBooNe experiment both claim an independent 3.8 $\sigma$ effect from standard
       neutrino physics. The LNSD signal with anti-neutrino
       oscillations from an accelerator would imply an additional mass-squared
       difference largely in excess of the Standard Model's values. The LSND signal
       ($87.9 \pm 22.4 \pm 6.0$) represents a 3.8 $\sigma$ effect at L/E values of 
       about $0.5 \div 1.0$ meter/MeV. The recent MiniBooNe result, confirming the LNSD result,
       indicates a neutrino oscillation signal both in neutrino and antineutrino with
       $\Delta m_{new}^2\sim$ 0.01 to 1.0 eV$^2$.

   \item  The apparent {\em disappearance signal} in the $\overline{\nu}_e$ events 
       detected from {\bf (a)} near-by nuclear reactors and {\bf (b)} from Mega-Curie k-capture calibration
       sources in the solar $\nu_e$ Gallium experiments~\cite{larnessie_4}. \\
       {\bf (a)} Recently a re-evaluation 
       of all the reactor antineutrino spectra has increased the flux by about 4\% and a new 
       value of the neutron lifetime has been reported~\cite{reattori}. With such a new flux evaluations, 
       the ratio between the observed and predicted rates decreased to $0.927 \pm 0.023$,
       leading to a deviation of 3.0 $\sigma$ from unity (99.6 \% confidence level). 
       Reactor experiments however generally explore distances which are far away from the 
       perspective oscillatory region with $\Delta m_{new}^2\simeq 2$ eV$^2$, with
       perhaps the exception of the ILL experiment (at $\sim$ 9 m from the source) which had
       unfortunately a very modest statistical impact (68\% confidence level). \\
       {\bf (b)} SAGE and
       GALLEX experiments recorded the calibration signal produced by intense artificial 
       k-capture sources of $^{51}$Cr and $^{37}$Ar. The averaged ratio between the 
       source detected and predicted neutrino rates are  consistent with each other, 
       giving ($0.86 \pm 0.05$), about 2.7 $\sigma$ from unity and a broad range of values
       centered around $\Delta m_{new}^2\sim 2$ eV$^2$ and $\sin^2 2\theta_{new}\sim 0.3$.\\
       By combining the Gallum and the reactor anomalies the non oscillation hypothesis is disfavored at 3.6 $\sigma$.
       
   \item The existence of additional neutrino states may be also hinted -- or at least not
  excluded -- by cosmological data~\cite{wmap}.

\end{itemize}

       All recalled ``anomalies'' which have accumulated an impressive number of
standard deviations, may indeed hint at a unified scheme in which the values of
$\Delta m_{new}^2$ may have a common origin, the different values of $\sin^2 2 \theta_{new}$
for different channels reflecting the structure of $U_{(4,k)}$ matrix 
(or even with an higher number of neutrinos) with $ k = \mu$ and $e$.

       The ultimate goal is to prove the existence of additional neutrino states and the corresponding oscillation parameters.
The proposed CERN experiment may be in a unique position to be
able to investigate such oscillation scenario with high sensitivity. The direct,
unambiguous measurement of an oscillation pattern requires necessarily the
(simultaneous) observation at least at two different positions. It is only in this way that
the new values of $\Delta m^2$ and of $\sin^2 2\theta$ can be separately identified. All other
experimental non accelerator programmes under investigation will not focus with an
equal sensitivity to the direct parameters of the oscillation phenomenon.

        The CERN experiment with the novel development of a large mass LAr-TPC
--~routinely operated at CNGS over the last 2.5 years and to be moved to CERN --
complemented with magnetic spectrometers for the charge determination introduces
 new relevant features, {\em allowing a simultaneous clarification of all the
above described ``anomalies''}. More precisely we will provide:

 \begin{itemize}
 
      \item L/E oscillation path-lengths appropriate to match to the $\Delta m^2$ 
            window for the expected anomalies;
	    
      \item ``imaging'' detector capable to identify unambiguously \underline{all} reaction channels
       with a {\em ``Gargamelle class''} LAr-TPC;
       
      \item magnetic spectrometers able to determine muon charge with few \% mis-identification and momentum in a broad range;
      
      \item   clean measure of interchangeable $\nu$ and anti-$\nu$ focused beams;
      
      \item   very high event rates due to large detector masses, allowing to record relevant
       effects at the \% level ($> 10^6 \nu_\mu$ and $\simeq 10^4 \nu_e$);
       
      \item   initial $\nu_e$ and $\nu_\mu$ components cleanly identified.
  \end{itemize}

        A similar LAr-TPC accelerator experiment, MicroBooNe, has been
approved at FermiLab. The experiment (widely based on ICARUS experience)
should start data taking around end 2014 with one single site at about 470 m from the
target and $\simeq~60$~ton fiducial mass. The average neutrino beam is about 0.8 GeV and
$6 \times 10^{20}$~pot in neutrino mode from the 8 GeV Booster (2-3 year run). The expected
neutrino signal events at the LSND best fit ($\Delta m^2 =1.2$ eV$^2$, $\sin^2 2\theta =0.003$
and 475-1250 MeV) are $\sim~$ 70 events with an expected background of 150 events. There
is apparently no immediate plan to run with anti-neutrinos, since event rates
will be even lower (by a factor of 3?).

        The FermiLab experiment should be compared to the present proposal whose
         $\nu_e$ background after 2 years of neutrino run, $9.0 \times 10^{19}$ pot
and $E_\nu < 6$ GeV, is about  6150 events. The expected signal for 
$\Delta m^2 =2.0$ eV$^2$, and the smaller value of $\sin^2 2\theta =0.002$,
 is $\sim 1450$ events. The huge differences in rates are mainly associated with the much greater 
detector mass of the CERN proposal, its simultaneous detection in two or more positions and to the
higher energy of the CERN neutrino beam. Moreover in our proposal spectrometers allow 
$\nu_{\mu}$ disappearance search  thereby filling in the project physics reach
and to constrain  systematic errors at few \%level.

       A Double Liquid Argon TPC ``proposal'' has also been visualized for
FermiLab with a second 1 kton LAr detector to be constructed, but with no muon
spectrometer at least to our knowledge.

\subsubsection*{The experimental proposal}

We report here on the experimental proposal~\cite{larnessie_1} currently under scrutiny by CERN 
committees.
The experiment follows the setting up of a new neutrino beam at SPS in a short time schedule. 
We deem mandatory that both beam and experiment be ready by December 2015, in order 
to be competitive with the expected flow of neutrino physics results in the international landscape.

\begin{figure}[htbp]
\begin{center}
  \includegraphics[width=0.9\textwidth]{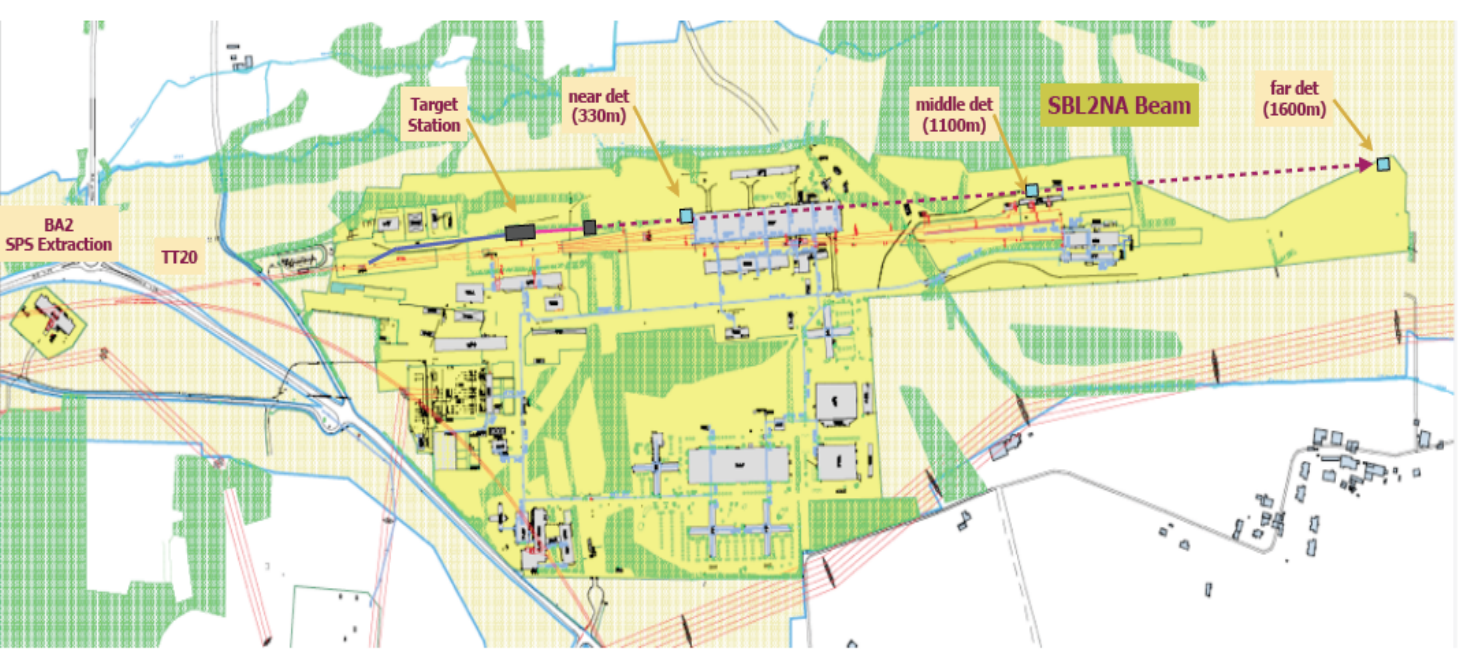}
    \caption{The new SPS North Area neutrino beam layout. Main parameters are: primary beam: 100~GeV; fast extracted from SPS; 
    target station next to TCC2, $\sim$11 m underground; decay pipe: 100 m, 3 m diameter; beam dump : 15 m of Fe with graphite core, 
    followed by muon stations; neutrino beam angle: pointing upwards; at $\sim$3 m in the far detector $\sim$5 mrad slope.}
    \label{larnessie_fig1}
\end{center}
\end{figure}

\begin{figure}[htbp]
\begin{center}
  \includegraphics[width=0.8\textwidth]{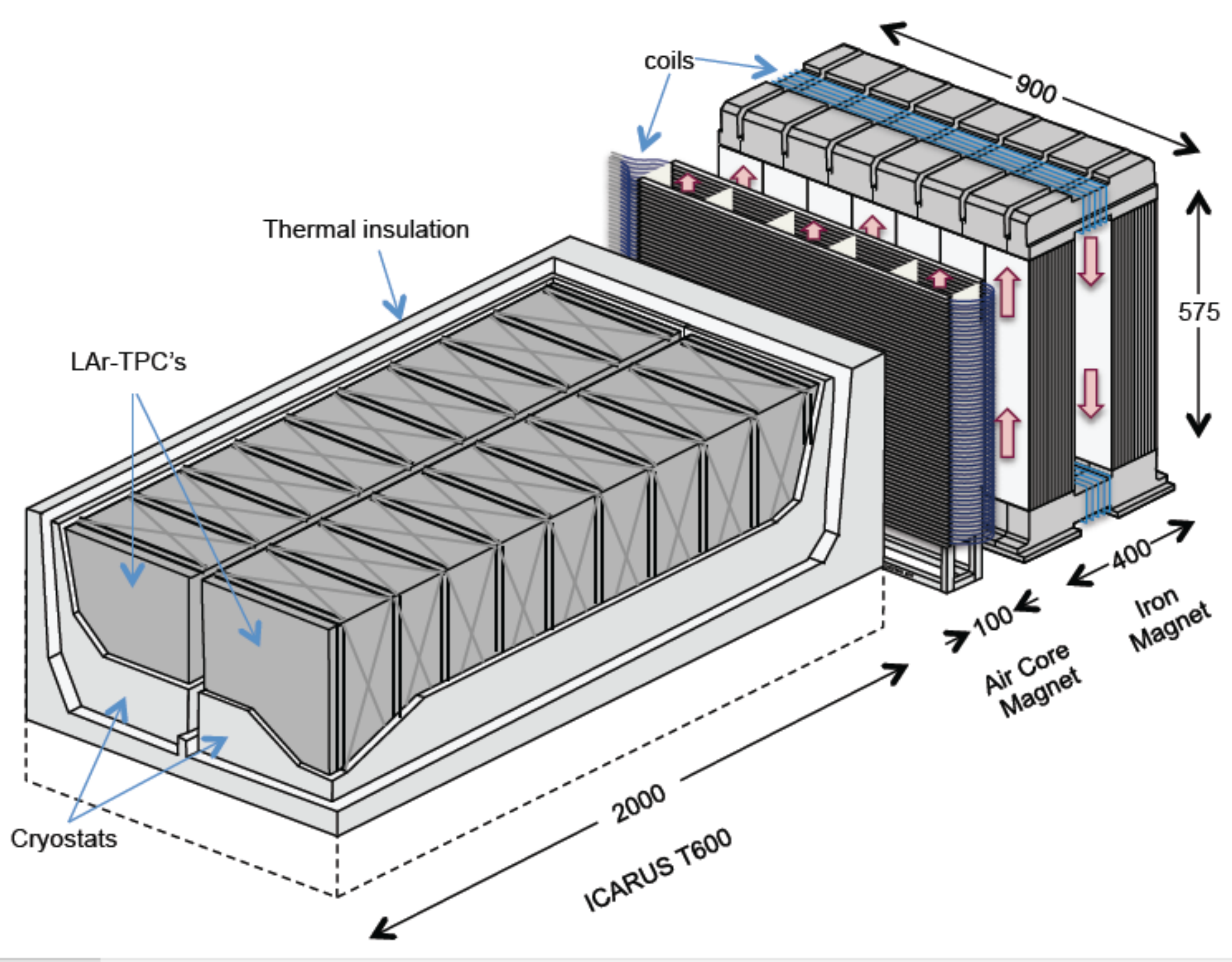}
    \caption{3D sketch of the ICARUS/NESSiE detector layout at the far site.}
    \label{icarus_nessie_far4}
\end{center}
\end{figure}

The experiment is based on two identical LAr-TPCÕs~\cite{larnessie_2} complemented by magnetized 
spectrometers~\cite{larnessie_3} detecting electron and muon neutrino events at Far and Near
positions, 1600 m and 300 m from the proton target, respectively 
(Figure~\ref{larnessie_fig1}). The project will exploit the ICARUS T600 detector, the largest 
LAr-TPC ever built with a size of about 600 ton of imaging mass, now running in the LNGS underground 
laboratory exposed to the CNGS beam, moved at the CERN ÒFarÓ position. 
An additional 1/4 of the T600 detector (T150) will be constructed and located in the Near
position. Two spectrometers will be placed 
downstream of the two LAr-TPC detectors to greatly complement the physics capabilities.
Spectrometers will exploit a classical dipole magnetic field with iron slabs, and a new concept 
air-magnet, to perform charge identification and muon momentum measurements 
from sub-GeV to several GeV energy range, over a transverse area larger than 50 m$^2$.
A 3D sketch of the detector layout at the far site is shown in Figure~\ref{icarus_nessie_far4}. 

At the two positions the energy spectra of the $\nu_e$ beam component
must coincide but for second order effects which can be reliably reproduced.
In absence of oscillations, since all cross sections and experimental biases cancel out, the observed event distributions
at the near and far detectors must be the same.
Any difference can be ascribed to the possible existence of 
$\nu$-oscillations due to additional neutrinos with new mixing angles $\sin^2 2\theta_{ij}$ and mass differences 
$\Delta m_{ij}^2$ larger than those measured in the standard three neutrino family scheme.

The superior quality of the LAr imaging TPC, now widely experimentally demonstrated, and in particular its unique {\em electron} - $\pi^0$ 
discrimination allows full rejection of backgrounds and offers a lossless $\nu_e$ detection capability. The determination of the muon charge 
with the spectrometers allows the full separation of $\nu_{\mu}$ from $\overline{\nu}_\mu$ and an improved control of systematics
from muon mis-identification.

Two main anomalies will be explored with both neutrino and anti-neutrino focused beams. The first anomaly, emerged in 
radioactive sources and reactor neutrino experiments~\cite{larnessie_4}, could originate from
 $\nu_e$ ($\overline{\nu}_e$) and/or of the
$\nu_\mu$ ($\overline{\nu}_\mu$) converted into ``invisible'' (sterile) 
components, leading to observation of oscillatory, 
distance dependent, disappearance rates. In a second anomaly (following LSND and MiniBooNE observations~\cite{larnessie_5}) 
some distance dependent $\nu_\mu \rightarrow \nu_e$ oscillations may be observed as a $\nu_e$ excess, 
especially in the antineutrino channel. The disentangling of $\nu_{\mu}$ from $\overline{\nu}_\mu$ will allow to exploit the interplay 
of the different possible oscillation scenarios, as well as the interplay between disappearance and appearance of different neutrino 
states and flavors. Moreover the NC/CC ratio will provide a sterile neutrino oscillation signal by itself and it will beautifully complement
the normalization and the systematics studies. This experiment will offer remarkable discovery potentialities, collecting a very large 
number of unbiased events both in the neutrino and antineutrino channels, largely adequate to definitely settle the origin of the 
$\nu$-related anomalies.

\subsubsection*{The new SPS neutrino facility}

\begin{figure}[htbp]
\begin{center}
  \includegraphics[width=0.45\textwidth]{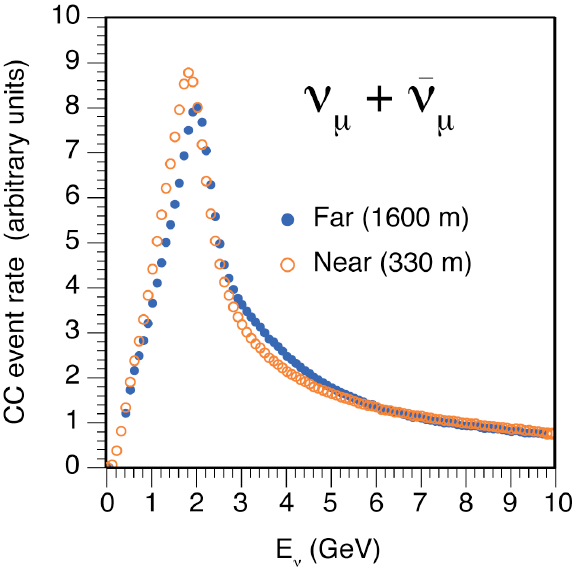}
  \includegraphics[width=0.45\textwidth]{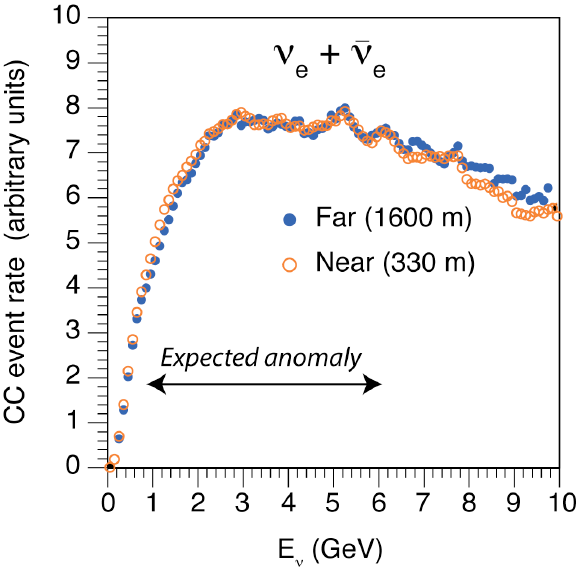}
    \caption{Muon (left) and electron (right) neutrino CC interaction spectra, at near and far positions.}
    \label{larnessie_fig2}
\end{center}
\end{figure}

To explore the interesting neutrino $\Delta m^2 \sim 1$ eV$^2$ region the ÒFarÓ distance has been chosen at 1.6 km with a central 
value of the on-axis neutrino beam energy spectrum around $E_{\nu} \sim$ 2 GeV (Figure~\ref{larnessie_fig2}). A proton beam 
intensity of $4.5\times 10^{19}$ pot/year at 100 GeV energy has been assumed as a conservative reference in order to produce 
high intensity $\nu$ beam and to minimize the beam related background expected at the near detector located at 300 m. 
A fast proton extraction from the SPS is also required for the LAr-TPC operation at surface in order to effectively separate the beam related events 
among the cosmic ray background.

\subsubsection*{Expected sensitivities to neutrino oscillations}

A complete discussion of $\nu_\mu \rightarrow \nu_e$ oscillation search both in appearance and disappearance modes is
presented in the SPSC-P345 document~\cite{larnessie_2}, that includes the genuine event selection and background rejection in the 
LAr-TPC. In particular, due to the excellent $\pi^0$ to electron separation, a $\pi^0$ rejection at $10^3$ level is obtained when 
requiring at least 90\% electron recognition efficiency. The effects of the high-energy event tail in the event selection was carefully 
studied: the resulting background is negligible, of the same order of the residual NC background.

In addition to the $\nu_\mu \rightarrow \nu_e$ oscillation searches, $\nu_{\mu}$ oscillation studies in disappearance mode 
are discussed at length in the SPSC-P-343 proposal~\cite{larnessie_3}, by using large mass spectrometers with high capabilities in charge identification and 
muon momentum measurement. It is important to note that all sterile neutrino models predict large $\nu_{\mu}$ disappearance effects together 
with $\nu_e$ appearance/disappearance. To fully constrain the oscillation searches, the $\nu_{\mu}$ disappearance mode has to has 
be carefully investigated. Much higher disappearance probabilities (with relative amplitudes as large as 10\%) are expected than in 
appearance mode. 
The spectrometers will allow to correctly identify about 40\% of all the CC events produced in, and escaped 
from, the LAr-TPCs, both at the near and far sites. That will greatly increase the fraction of CC events with charge identification and 
momentum measurement. Therefore a complete measurement of the CC event spectra will be possible, along with the measurement of 
the NC/CC event ratio 
(in synergy with the LAr-TPC), and the associated background systematics.

The large mass of the magnets ($\sim 3$ kton) will allow an internal check of the NC/CC ratio in an extended energy range, and an independent 
measure of the CC oscillated events.

\begin{figure}[htbp]
\begin{center}
  \includegraphics[width=0.8\textwidth]{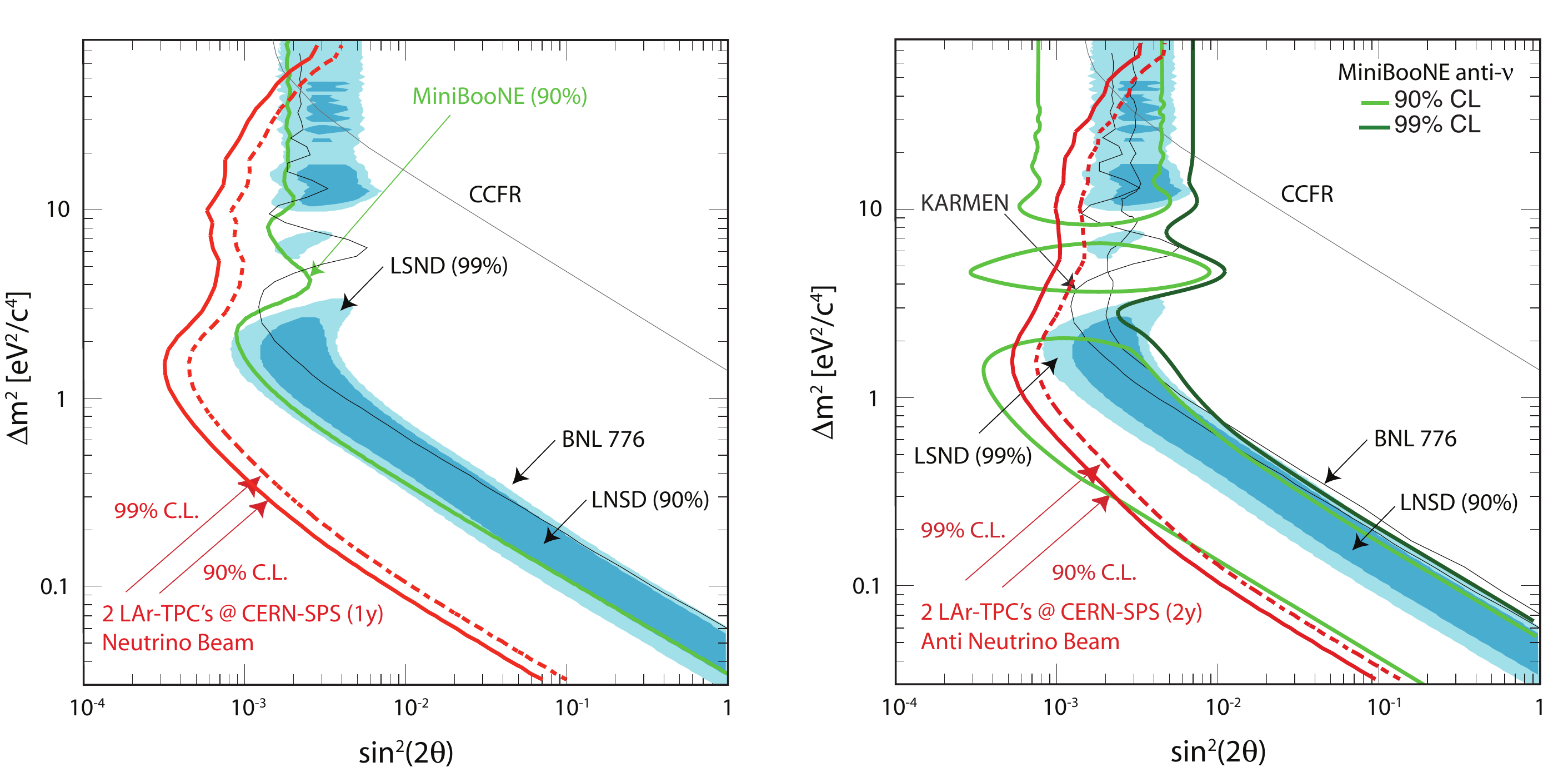}
    \caption{Expected sensitivity for the proposed experiment exposed at the CERN-SPS neutrino beam (left) and anti-neutrino (right) 
    for $4.5\times 10^{19}$ pot (1 year) and $9.0\times 10^{19}$ pot (2 years), respectively. The LSND allowed region is fully explored 
    in both cases.}
    \label{larnessie_fig3}
    \end{center}
\end{figure}

We are sensitive to $\sin^2 2\theta$ down to $3\times 10^{-4}$ (for $|\Delta m^2|  > 1.5$ eV$^2$) and $|\Delta m^2|$ down to  $0.01$ 
eV$^2$ (for $\sin^2 2\theta = 1$) at 90\% C.L. for the $\nu_\mu \rightarrow \nu_e$ transition with one year exposure 
($4.5\times 10^{19}$ pot) at the CERN-SPS $\nu_{\mu}$ beam (Figure~\ref{larnessie_fig3} left). The parameter space region allowed 
by the LSND experiment is fully covered, except for the highest $\Delta m^2$ region. The sensitivity has been computed according to 
the above described particle identification efficiency and assuming a 3\% systematic uncertainty in the prediction of ÒFarÓ to ÒNearÓ 
$\nu_e$ ratio. A further control of the overall systematics will be provided by the LAr and spectrometer combined measurement of 
CC spectra in the near site and over the full energy range.

In anti-neutrino focusing, twice as much exposure ($0.9\times 10^{20}$ pot) allows to cover both the LSND region and the new 
MiniBooNE results (Figure~\ref{larnessie_fig3} right)~\cite{larnessie_5}. Both favoured MiniBooNE parameter sets, corresponding to 
two different energy regions in the MiniBooNE antineutrino analysis, fall well within the reach of this proposal.

\begin{figure}[htbp]
\begin{center}
 \includegraphics[width=0.4\textwidth]{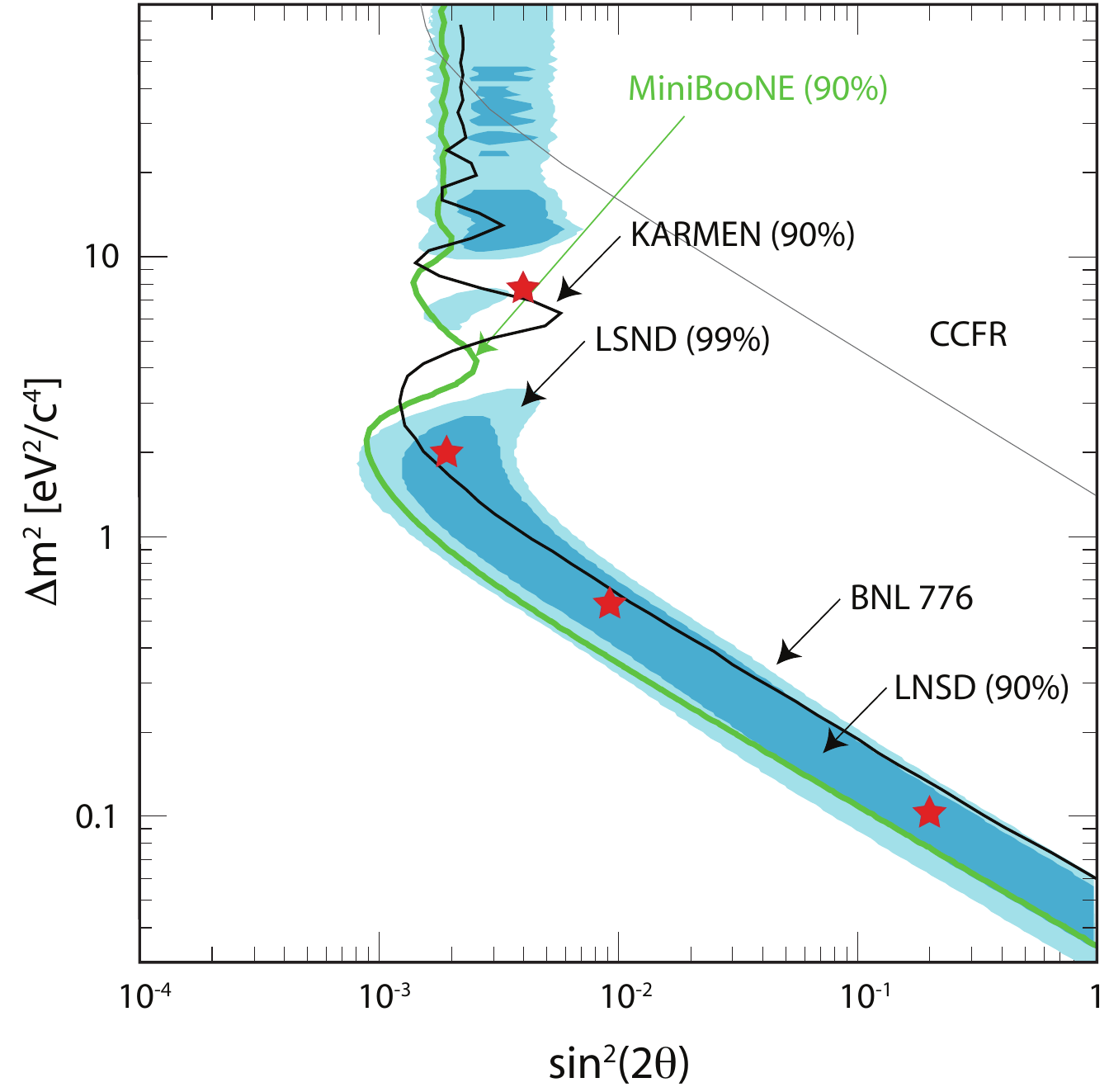}
  \includegraphics[width=0.55\textwidth]{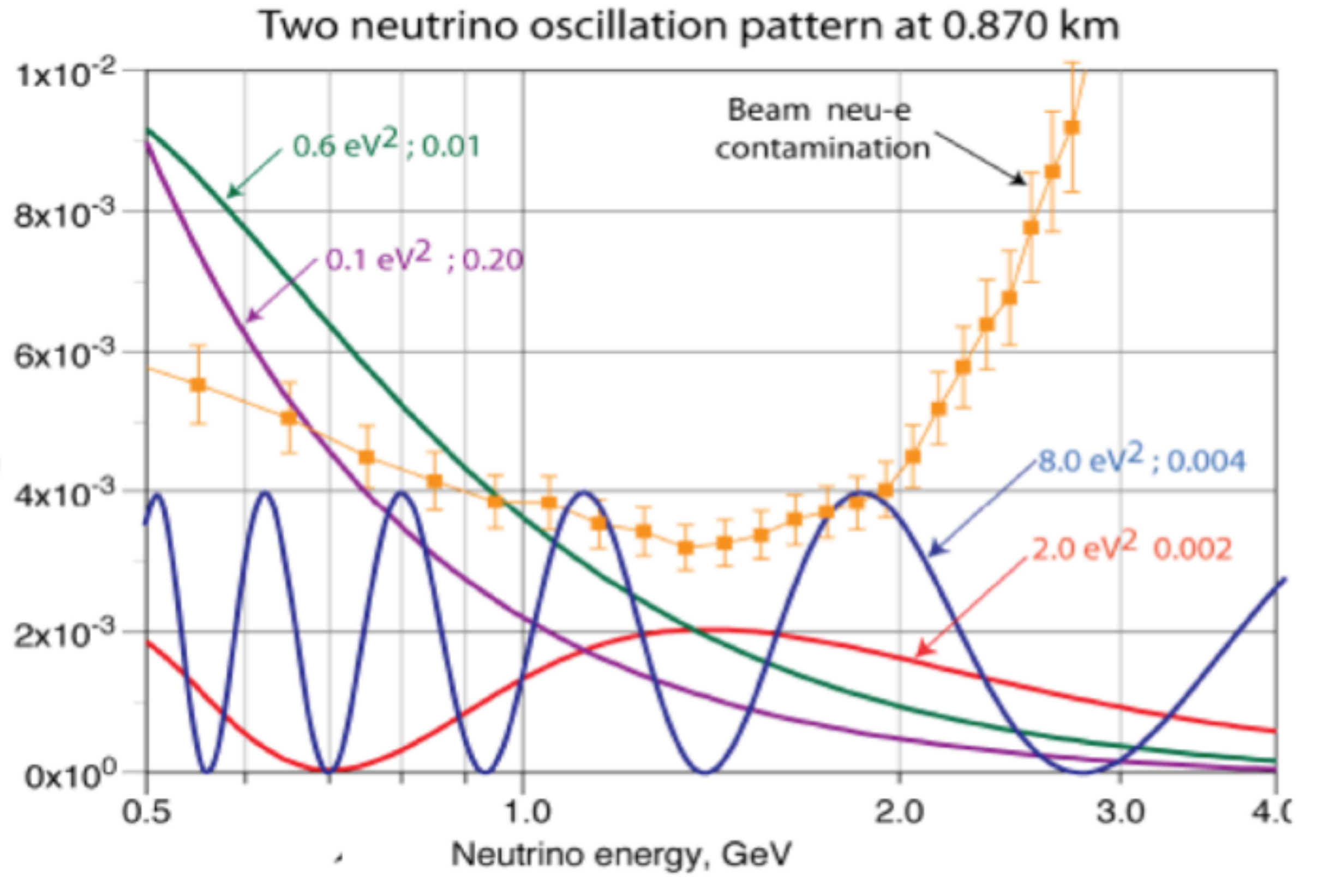}
     \caption{The present method, unlike LSND and MiniBooNE, determines both the mass difference and the value of the mixing angle.
     Very different and clearly distinguishable patterns ($\star$, 1--4) are possible, depending on the values in the  
     ($\Delta m^2$  -- $\sin^2 2\theta$) plane. The intrinsic $\nu_e$ background (5) is also shown.}
    \label{larnessie_fig3bis}
    \end{center}
\end{figure}

    It should be remarked that the observation of the $\nu_e$ energy distribution at two different
distances offers the possibility of distinguishing both the mass difference  $\Delta m^2$ and the
mixing angle $\sin^2 2 \theta$ independently. In Figure~\ref{larnessie_fig3bis} some different oscillation parameter
values are identified by the electron neutrino spectrum which appears to be extremely sensitive to
the actual values of LSND like $\nu_\mu \rightarrow \nu_e$  sterile neutrino events. 
The presence of the intrinsic $\nu_e$ beam associated background is also
shown.

\begin{figure}[htbp]
\begin{center}
  \includegraphics[width=0.7\textwidth]{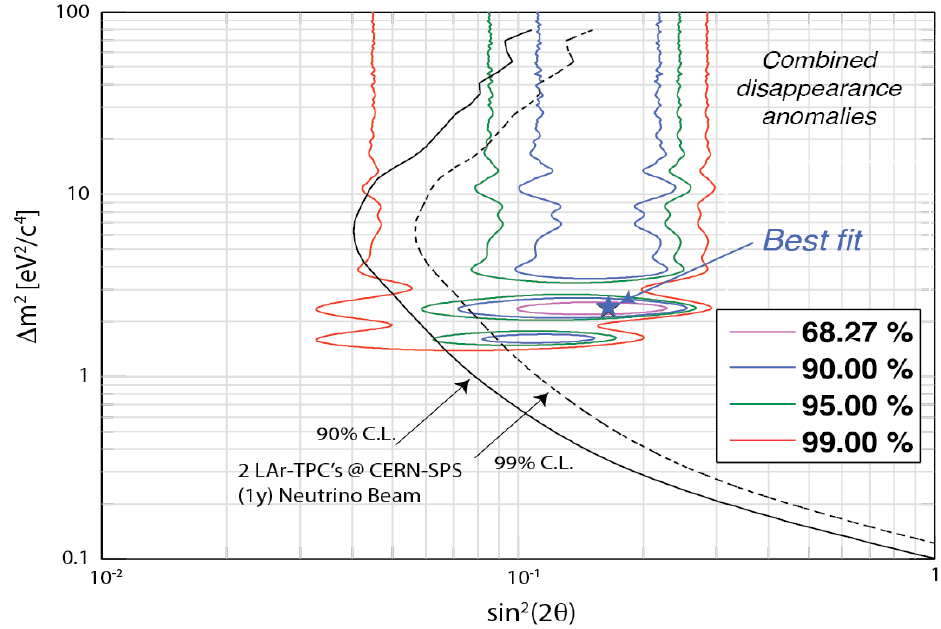}
     \caption{Oscillation sensitivity in $\sin^2 2\theta$ vs $\Delta m^2$ distribution for 1 year data taking. 
     A 3\% systematic uncertainty on energy spectrum is included. Combined ÒanomaliesÓ from reactor neutrino, 
     Gallex and Sage experiments are also shown.}
    \label{larnessie_fig4}
    \end{center}
\end{figure}

In Figure~\ref{larnessie_fig4} the sensitivity for $\nu_e$ disappearance search in the $\sin^2 2\theta$, $\Delta m^2$ 
plane is shown for one year data taking. The oscillation parameter region related to the anomalies from the combination 
of the published 
reactor neutrino experiments, GALLEX and SAGE calibration sources experiments~\cite{larnessie_4} is completely explored.

\begin{figure}[htbp]
\begin{center}
  \includegraphics[width=0.7\textwidth]{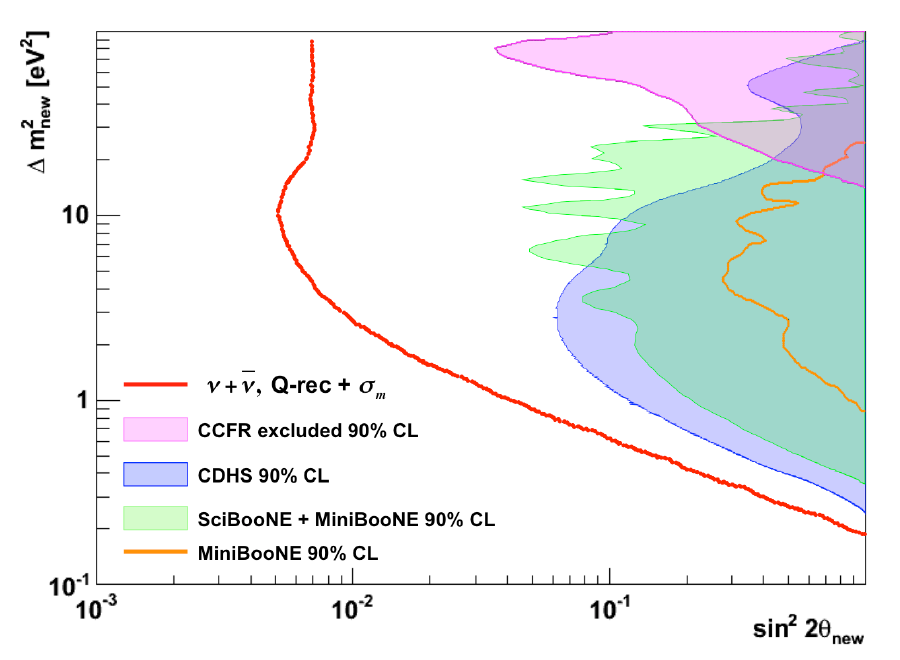}
     \caption{Sensitivity plot (at 90\% C.L.) considering 3 years of the CERN-SPS beam (2 years in anti-neutrino and 1 year in neutrino mode) 
     from CC events fully reconstructed in NESSiE$+$LAr. Red line: $\nu_{\mu}$ exclusion limit. The three filled areas correspond to the present 
     exclusion limits on the $\nu_{\mu}$ from CCFR, CDHS and SciBooNE+MiniBooNE experiments (at 90\% C.L.). 
     Orange line: recent exclusion limits on $\nu_{\mu}$ from MiniBooNE alone measurement.}
    \label{larnessie_fig5}
    \end{center}
\end{figure}

The $\nu_{\mu}$ disappearance signal is well studied by the spectrometers, with a very large statistics, and disentangling of $\nu_{\mu}$ and 
$\overline{\nu}_\mu$  interplay~\cite{larnessie_6}.  Figure~\ref{larnessie_fig5} shows the sensitivity plot (at 90\% C.L.) for two 
years negative-focusing plus one year positive-focusing. A large extension of the present limits for $\nu_{\mu}$ by CDHS and the 
recent SciBooNE+MiniBooNE will be achievable in $\sin^2 2\theta,\; \Delta m^2$.

For one year of operation, either with negative or positive polarity beam, Table~\ref{larnessie_tab1} reports the expected interaction rates in the 
LAr-TPCs at the near (fiducial 119 ton) and far locations (fiducial 476 ton), and the expected rates of fully reconstructed events in the NESSiE 
spectrometers at the near (fiducial 241 ton) and far locations (fiducial 661 ton), with and without LAr contribution. Both $\nu_e$ and $\nu_{\mu}$ 
disappearance modes will be used to add conclusive  information on the sterile mixing angles as shown in the Table ~\ref{larnessie_tab2}.

\begin{table}[htbp!]
\footnotesize
  \begin{center}
  \begin{tabular}{lcccc}
  \hline
  & NEAR                            & NEAR             & FAR                        & FAR            \\
  &       (Negative foc.)           & (Positive foc.)  & (Negative foc.)            & (Positive foc.) \\
  \hline
 $\nu_e + \overline{\nu}_e$ (LAr) & 35 K & 54 K & 4.2 K & 6.4 K \\
 $\nu_{\mu} +  \overline{\nu}_{\mu}$ (LAr) & 2000 K & 5200 K & 270 K & 670 K \\ 
 Appearance Test Point & 590 & 1900 & 360 & 910 \\
                                            &        &         &      &       \\
 $\nu_{\mu}$ CC (NESSiE$+$LAr) & 230 K & 1200 K & 21 K & 110 K \\
  $\nu_{\mu}$ CC (NESSiE alone) & 1150 K & 3600 K & 94 K & 280 K \\
  $\overline{\nu}_{\mu}$ CC (NESSiE$+$LAr) & 370 K & 56 K & 33 K & 6.9 K \\
  $\overline{\nu}_{\mu}$ CC (NESSiE alone) & 1100 K & 300 K & 89 K & 22 K \\
 Disappearance Test Point & 1800 & 4700 & 1700 & 5000 \\
 \hline
  \end{tabular}
  \caption{The expected rates of interaction (LAr) and reconstructed (NESSiE) events for 1 year of operation.Values for 
  $\Delta m^2$ around 2 eV$^2$ are reported as example.}
  \label{larnessie_tab1}
  \end{center}
\end{table}

\begin{table}[htbp!]
\footnotesize
  \begin{center}
  \begin{tabular}{cll}
  \hline
  Osc. type & Neutrinos & Experiments \\
  \hline
  $\theta_{12}$ & $\nu_e$ (solar, reactors) & SNO, SK, Borexino, Kamland \\
  $\theta_{23}$ & $\nu_{\mu}$ (atmospheric,accelerators) & SK, Minos, T2K \\
  $\theta_{13}$ & $\nu_e$ (reactors) & Daya Bay, Reno, Double Chooz \\
   $\theta_{14}$ & $\nu_e$ (reactors, radioactive sources) & SBL Reactors, Gallex, Sage. {\bf This Proposal} \\
  $\theta_{24}$ & $\nu_{\mu}$ (accelerators) & CDHS, MiniBooNE. {\bf This Proposal} \\
  \hline
  \end{tabular}
  \caption{Measurements of the mixing angle as provided by different experiments.}
  \label{larnessie_tab2}
  \end{center}
\end{table}

\end{document}